\documentclass[authoryear,11pt]{article}
\usepackage[colorlinks,bookmarksopen,bookmarksnumbered,citecolor=red,urlcolor=red]{hyperref}
\usepackage{graphicx}
\usepackage[authoryear]{natbib}
\usepackage{amsmath,psfrag,epstopdf,amssymb,psfrag}
\usepackage{multirow}
\usepackage{setspace}
\newcommand{\X}{\mathbf{X}}

\newcommand{\Z}{\mathbf{Z}}
\newcommand{\y}{\mathbf{y}}

\newcommand{\T}{\scriptscriptstyle \mathrm{T}}

\psfrag{n=10, J=20}{\scriptsize $n=10,\; J=20$}
\psfrag{n=40, J=80}{\scriptsize $n=40,\; J=80$}
\psfrag{Density}{\scriptsize density}
\def\frac#1#2{{#1\over#2}}

\begin{document}

\noindent
\textsc{The wild bootstrap for multilevel models}%
\vskip 5mm
\noindent Lucia Modugno and Simone Giannerini
\vskip 2mm
\noindent Department of Statistical Sciences

\noindent University of Bologna

\noindent via Belle Arti, 41 - 40126 Bologna, Italy

\noindent lucia.modugno@unibo.it

\noindent simone.giannerini@unibo.it

\vskip 4mm
\baselineskip=18pt

\par
\begin{abstract}
    In this paper we study the performance of the most popular bootstrap schemes for multilevel data. Also, we propose a modified version of the wild bootstrap procedure for hierarchical data structures. The wild bootstrap does not require homoscedasticity or assumptions on the distribution of the error processes. Hence, it is a valuable tool for robust inference in a multilevel framework. We assess the finite size performances of the schemes through a Monte Carlo study. The results show that for big sample sizes it always pays off to adopt an agnostic approach as the wild bootstrap outperforms other techniques.
\end{abstract}

\section{Introduction}\label{sec:intro}
\emph{Multilevel data} consist of units of analysis of different type which are hierarchically clustered. In a strictly nested data structure, the term \emph{levels} represents the different types of unit of analysis, i.e.~the various types of grouping; in particular, the most detailed level is called the first (or the lowest) level.
\par
A meaningful example of multilevel data comes from studies on educational achievement, in which pupils, teachers, classrooms, schools, district, and so on, are clustered one within the other, and they might all be units of analysis, each described by own variables. Hierarchical data often occur also in social sciences: economists and political scientists frequently work with data measured at multiple levels in which individuals are nested in geographic divisions, institutions or groups, and so forth. Furthermore, other particular structures of data can be thought as multilevel: the repeated measurements over time on an individual, the respondents to the same interviewer and also subjects within a particular study among those of a meta-analysis can be considered groups of observations, and, consequently, be treated as multilevel data.
\par
The idea behind modelling multilevel data is that living environments affect individual behaviours, and contextual effects are due to social interactions within an environment. In general, individuals can influence and be influenced by various type of contexts: spatial, temporal, organizational and socio-economic-cultural. As~\citet{Kreft} put it, ``the more individuals share common experiences due to closeness in space and/or in time, the more they are similar, or, to a certain extent, duplications of each other"; in other words, performances of pupils in the same classroom tend to be more similar than those from a different classroom because of sharing contexts. This is one of the reasons why the specificity of multilevel data cannot be ignored, because the observations within one group are not independent of each other, as traditional models  require. If standard statistical analyses, which generally assume independent observations, are performed on multilevel data (the so-called \emph{naive pooling}
strategy), results may be misleading.
\par
Inference in multilevel models usually relies upon maximum likelihood methods (see for example \citet{Skrondal_book}, \citet{Raudenbush} and \citet{Searle}) that mostly use asymptotic approximations for the construction of test statistics and estimation of variances. If the sample size is not large enough, the asymptotic approximation does not hold and can lead to incorrect inferences. By using bootstrap methods, under some regularity conditions, it is possible to obtain a more accurate approximation of the distribution of the statistics. The original bootstrap procedure has been studied in detail by \citet{Efron} for independent and identically distributed (i.i.d.) observations. An extensive discussion of bootstrap methods for a variety of statistical models and for different data structures can be found in \citet{Davison}, and in particular for the multilevel structure in \citet{Leeden} and \citet{Goldstein2010}. In the case of hierarchical data three general bootstrap approaches are available and well established: the parametric
bootstrap, the residual bootstrap and the cases bootstrap.
\par
The aim of the paper is twofold: first, we review and test the finite size performance of the three bootstrap schemes for multilevel data by means of a simulation study; second, we propose a wild bootstrap procedure for multilevel data. The wild bootstrap does not assume homoscedasticity and, for this reason, can reveal appropriate for inference robust to heteroscedasticity of unknown form. The paper is organized
as follows. Section \ref{sec:boot} presents the three bootstrap schemes for multilevel models. In Section \ref{sec:wild} we introduce the wild bootstrap procedure for the linear regression model and extend it to the case of hierarchical data. Section \ref{sec:MC} presents a Monte Carlo study that compares all the methods under different scenarios. Finally, Section \ref{sec:concl} contains the conclusions.

\section{Bootstrap procedures for the multilevel model}\label{sec:boot}
 Resampling schemes for multilevel models have to take into account the hierarchical structure of data. Hence, the classic bootstrap procedures need to be adapted. The main bootstrap approaches for a multilevel model are discussed for example in~\citet{Leeden} and \citet{Goldstein2010}):
\begin{enumerate}
  \item the parametric bootstrap (resamples from the fitted distribution of the error processes)
  \item the residual bootstrap (resamples residuals from the fitted model)
  \item the cases bootstrap (resamples entire cases)
\end{enumerate}
The schemes differ in the underlying assumptions. We illustrate them by considering the following two-level model that includes $k$ level-1 covariates with fixed coefficients ($\mathbf{x}_{ij}$) and $p$ covariates with random coefficients ($\mathbf{z}_{ij}$)
\begin{equation}\label{eq:general model}
y_{ij}= \mathbf{x}_{ij}^{\T}\boldsymbol{\beta}+\mathbf{z}_{ij}^{\T}\mathbf{u}_j+\epsilon_{ij}, \quad \text{for}\quad i=1,\ldots,n_j\quad \text{and for}   \quad j=1,\ldots,J;
\end{equation}
with $\epsilon_{ij} \sim \text{NID}(0,\sigma^2_{\epsilon})$, $\mathbf{z}_{ij}^{\T}=\left[\begin{array}{cccc} 1&z_{1j}&\ldots&z_{pj}\end{array}\right]$ and $\mathbf{u}_j=\left[\begin{array}{cccc} u_{0j}&u_{1j}&\ldots&u_{pj}\end{array}\right]^{\T} \sim \text{NID}\big(\mathbf{0},\boldsymbol{\Sigma}\big)$, where the variance-covariance matrix of the random effects has the following form
\[
\boldsymbol{\Sigma}=\left[\begin{array}{cccc}\sigma^2_{u0} & \sigma_{u01} & \ldots & \sigma_{u0p}\\
\sigma_{u10} & \sigma^2_{u1} & \ldots & \sigma_{u1p}\\
\vdots & \vdots& \vdots& \vdots\\
\sigma_{up0} & \sigma_{up1}& \ldots & \sigma^2_{up} \\
\end{array}\right].
\]
Moreover, it is assumed that the random effects for the group j, the vectors $\mathbf{u}_j$, and the within-group errors, $\epsilon_{ij}$, are independent. Finally, we denote by
$\boldsymbol{\hat{\theta}}=\{\boldsymbol{\hat{\beta}}, \hat{\sigma}^2_{\epsilon},\boldsymbol{\hat{\Sigma}}\}$ the (restricted) maximum likelihood estimates of the parameters of the model (\ref{eq:general model}).
\subsection{The parametric bootstrap}\label{subsec:parametric}
The parametric bootstrap assumes that the covariates are fixed and that both the model and the error distributions are correctly specified. Compared to the other two approaches, such scheme has the strongest requirements. In practice, the parametric bootstrap for model (\ref{eq:general model}) generates resamples as follows:
 \begin{enumerate}
   \item draw $N$ elements $\hat{e}_{ij}^*$ from  the estimated distribution of level-1 errors, $\hat{F}_{\epsilon}\sim N(0,\hat{\sigma}^2_{\epsilon})$, where $N=\sum_{j=1}^J n_j$ is the total sample size;
   \item draw  $J$ (p+1)-vectors of elements $\mathbf{\hat{u}}_{j}^*$ from the estimated distribution of the random effects, $\hat{F}_u\sim N\big(\mathbf{0},\boldsymbol{\hat{\Sigma}}\big)$;
     \item generate the bootstrap responses as $y_{ij}^*=\mathbf{x}_{ij}^{\T}\boldsymbol{\hat{\beta}}+\mathbf{z}_{ij}^{\T}\mathbf{\hat{u}}_j^*+\hat{e}_{ij}^* \quad \forall i,j;$
   \item compute the bootstrap value $\boldsymbol{\hat{\theta}}^*$ on the generated sample;
   \item repeat steps $1-4$ B times as to obtain B sets of bootstrap replications of the parameters.
 \end{enumerate}
As remarked above, the parametric bootstrap is not robust with respect to any deviation from the normality assumption on the error terms so that severe problems can occur in such cases.

\subsection{The residual bootstrap}\label{subsec:residual}

The residual bootstrap was introduced in the multilevel framework by \citet{Carpenter}; it treats the covariates as fixed and assumes that the model specification is correct. No distributional assumptions on error terms are required but only variance homogeneity among groups. In the classic linear regression framework, the scheme resamples with replacements from the residuals of the fit. The implementation of this procedure in a multilevel model leads to a distortion because the residuals are \emph{shrunken} towards zero so that the true variability of the residuals is not reproduced in the resamples \citep{Goldstein2010}. Therefore, it is necessary to reflate the \emph{shrunken} residuals. The procedure generates bootstrap samples as follows:
 \begin{enumerate}
   \item compute level-2 and level-1 \emph{shrunken} residuals from model (\ref{eq:general model}), respectively
      \[
      \hat{\mathbf{u}}_j=\boldsymbol{\hat{\Gamma}}\Z^{\T}_j\big(\Z\boldsymbol{\hat{\Gamma}}\Z^{\T} + \hat{\sigma}^2_{\epsilon}\mathbf{I}_N\big)^{-1}\big(\y_j-\X_j\boldsymbol{\hat{\beta}}\big)
      \]
      \[
      \mbox{and}\quad\hat{e}_{ij}=y_{ij} - \mathbf{x}_{ij}^{\T}\boldsymbol{\hat{\beta}}-\mathbf{z}_{ij}^{\T}\mathbf{\hat{u}}_j\quad \forall i,j,
      \]
      where $\boldsymbol{\hat{\Gamma}}=\text{diag}\big(\boldsymbol{\hat{\Sigma}},\boldsymbol{\hat{\Sigma}}, \ldots,\boldsymbol{\hat{\Sigma}}\big)$ is the ML estimate of the block-diagonal variance-covariance matrix of the whole random-effects matrix $\mathbf{U}=[\mathbf{u}_1,\mathbf{u}_2, \ldots, \mathbf{u}_J]^{\T}$; both level-1 and level-2 residuals must be centered, since in the multilevel model their mean is not zero;
   \item reflate the (centered) residuals, that is, consider a transformation $\tilde{\mathbf{U}}=\mathbf{\hat{U}}\mathbf{A}$ such that $\tilde{\mathbf{U}}^{\T}\tilde{\mathbf{U}}/J= \boldsymbol{\hat{\Gamma}}$. In other words, we need to achieve that estimates of the variance obtained by means of the \emph{shrunken} residuals equal the maximum likelihood estimates obtained from the model; otherwise, the procedure would lead to downward biased estimates of the variance parameters. For level-2 residuals we have $    \tilde{\mathbf{U}}^{\T}\tilde{\mathbf{U}}/J=\mathbf{A}^T\hat{\mathbf{U}}^{\T}\hat{\mathbf{U}}\mathbf{A}/J=\mathbf{A}^{\T}\mathbf{S}\mathbf{A} = \boldsymbol{\hat{\Gamma}};$
    one possible choice of $\mathbf{A}$ is $\mathbf{A}=\mathbf{L}_S^{-1}\mathbf{L}_{\Gamma}$ where $\mathbf{L}_S$ and $\mathbf{L}_{\Gamma}$ are the Cholesky decompositions of $\mathbf{S}$ and $\boldsymbol{\hat{\Gamma}}$ respectively. The same procedure is applied to level-1 residuals;
   \item draw with replacement $J$ vectors $\mathbf{\hat{u}}_{j}^*$  from the set of reflated level-2 residuals, $\{\tilde{\mathbf{u}}_j\}$ ;
   \item draw with replacement $N$ elements $\hat{e}_{ij}^*$ from the set of reflated level-1 residuals $\{\tilde{e}_{ij}\}$;
   \item generate the bootstrap responses as $y_{ij}^*=\mathbf{x}_{ij}^{\T}\boldsymbol{\hat{\beta}}+\mathbf{z}_{ij}^{\T}\mathbf{\hat{u}}_j^*+\hat{e}_{ij}^* \quad \forall i,j;$
   \item compute bootstrap values $\boldsymbol{\hat{\theta}}^*$ on the generated sample;
   \item repeat steps $3-6$ B times as to obtain B bootstrap replications of the parameters.
 \end{enumerate}

Since, it does not rely on any distributional assumptions, the residual bootstrap is robust with respect to non-normality of the error processes. By means of a simulation study on a two-level model with $\chi_1^2$ errors, \citet{Carpenter} show that the empirical coverage of bootstrap confidence intervals is better for the residual bootstrap than for the parametric bootstrap; also, they show that the most important improvements concern the estimates of random parameters.

\subsection{The cases bootstrap}\label{subsec:cases}
The cases bootstrap for the linear regression model was proposed in \citet{Freedman} under the name \emph{pairs bootstrap}.  Of the three bootstrap procedures for multilevel model reviewed here, the cases bootstrap has the least restrictive assumptions: it assumes that only the hierarchical dependency in the data is correctly specified and considers the covariates as random variables. The procedure resamples entire cases as follows:
\begin{enumerate}
  \item draw with replacement $J$ units from the set of numbers $\{1,2,\ldots,J\}$; any drawn index $j'=1,\ldots,J$ is associated to a whole level-2 unit $(\y_{j'},\X_{j'},\Z_{j'})$;
  \item for each selected level-2 unit $(\y_{j'},\X_{j'},\Z_{j'})$, $j'=1,\dots,J$, draw with replacement a bootstrap sample $(\y_{j'}^*,\X_{j'}^*,\Z_{j'}^*)$ of size $n_{j'}$;
   \item compute the bootstrap value $\boldsymbol{\hat{\theta}}^*$ on the generated sample;
   \item repeat steps $1-3$ B times as to obtain B bootstrap replications of the parameters.
\end{enumerate}
The scheme of the cases bootstrap described above resamples both level-1 and level-2 units. However, whether this makes sense depends on the nature of the data. There may be instances in which it is correct to resample only level-1 (or level-2) units. For instance, when level-2 units are individuals and level-1 units are repeated measures, it is appropriate to resample only level-2 units; then, for each level-2 unit drawn all the associated level-1 units enter the resample. On the contrary, if level-2 units are countries or time points and level-1 units are individuals, it is more appropriate to resample only level-1 units within countries (or time points). For further discussion on the matter see \citet{Leeden}. The cases bootstrap (appropriately implemented) provides consistent estimators under heteroscedasticity at the price of less efficient estimators than the parametric and the residual bootstrap \citep{Flachaire99,Davison}.

\section{The wild bootstrap}\label{sec:wild}
The wild bootstrap is a technique aimed to obtain consistent estimators for the covariance matrix of the coefficients of a regression model when the errors are heteroscedastic. It was developed by \citet{Liu} following suggestions in \citet{Wu} and \citet{Beran}. Further evidences and refinements are provided in \citet{Flachaire} and \citet{Davidson}. Consider the classic linear regression model $
y_i=\mathbf{x}_i^{\T}\boldsymbol{\beta} + \nu_i,$
for $i=1,\ldots,n$. The disturbances $\nu_i$ are assumed to be mutually independent and to have zero mean, but they are allowed to be heteroscedastic. Moreover, the covariates are assumed to be strictly exogenous.\\ In the homoscedastic case, the variance of the residuals is proportional to $1-h_i$, where $h_i=\mathbf{x}_i^{\T}(\X^{\T}\X)^{-1}\mathbf{x}_i$ is the $i$-th diagonal element of the orthogonal projection matrix $\mathbf{H}=\X(\X^{\T}\X)^{-1}\X^{T}$. This suggests to replace the heteroscedastic residuals $\hat{v}_i$ with \begin{equation}\label{eq:HC}\tilde{v}_i= \hat{v}_i/\sqrt{(1-h_i)} \quad \text{or} \quad \tilde{v}_i= \hat{v}_i/(1-h_i)\end{equation} in order to reduce the bias of the variance estimator, as we use to do in the homoscedastic case by using the unbiased OLS estimator. These are two of the \emph{Heteroskedasticity Consistent Covariance Matrix Estimator} (HCCME) forms considered in \citet{MacKinnon}, that \citet{Flachaire} refers to as HC$_2$ and   HC$_3$ respectively. We omit to mention the other forms of HCCME since, as is shown in  \citet{MacKinnon} and \citet{Chesher}, HC$_2$ and   HC$_3$ outperform the others in terms of power and size of the tests; instead the two forms in (\ref{eq:HC}) cannot be ranked, although in some simulation experiments HC$_3$ has shown the least distortion (see for example \citet{Davidson}).
\par
The wild bootstrap procedure tries to recover the unknown form of heteroscedasticity of the errors by means of the following bootstrap data-generating process:
\begin{equation}\label{eq:bootsample}
y_i^*=\mathbf{x}_i^{\T}\boldsymbol{\hat{\beta}} + \tilde{v}_iw_i
\end{equation}
where $\boldsymbol{\hat{\beta}}$ is the vector of estimated regression coefficient, $\tilde{v}_i$ is one of the variants of HCCME (such as those in (\ref{eq:HC})) where the residuals have been transformed and the $w_i$ (for $i=1,\ldots,n$) are mutually independent errors drawn from an auxiliary distribution with zero mean and unit variance. While \citet{Mammen} suggests the following  asymmetric two-point distribution for the $w_i$
\begin{equation}\label{eq:F1}
F_1: \quad w_i = \left\{\begin{array}{lll}
                 -(\sqrt{5}-1)/2 & \text{with probability} & p= (\sqrt{5}+1)/(2\sqrt{5})\\
                 (\sqrt{5}+1)/2 &  \text{with probability} & 1-p,
                 \end{array}\right.
\end{equation}
\citet{Liu} mentions, instead, the distribution
\begin{equation}\label{eq:F2}
F_2: \quad w_i = \left\{\begin{array}{lll}1&\text{with probability}& 0.5\\-1& \text{with probability}& 0.5.\end{array}\right.\end{equation}
Based on the evidence of the simulation study, \citet{Davidson} recommend the auxiliary distribution $F_2$ rather than other versions. For further discussions on this choice, see also ~\citet{Liu} and \citet{Belsley}.
\par
The wild bootstrap procedure is as follows:
 \begin{enumerate}
 \item draw $n$ independent values, $w_i$, for $i=1,\ldots,n$, from an auxiliary distribution with zero mean and unit variance such as $F_1$ and $F_2$;
 \item generate the bootstrap samples according to Eq. (\ref{eq:bootsample})
   \item compute the bootstrap value $\boldsymbol{\hat{\theta}}^*$ on the generated sample $\y^*$;
\item repeat steps $1-3$ B times as to obtain B bootstrap replications of the parameters.
 \end{enumerate}
Note that also the pairs bootstrap, called cases bootstrap in multilevel models (subsection \ref{subsec:cases}),  is used to overcome the problem of heteroscedasticity of unknown form. However, \citet{Flachaire} shows that the version of wild bootstrap with $F_2$ in (\ref{eq:F2}) recommended in \citet{Davidson}, provides better numerical performance in terms of false rejection probability and power of a test than both other versions of wild bootstrap and the pairs bootstrap. Given these results for regression models, we expect that the wild bootstrap behaves similarly when applied to multilevel data. However, to our knowledge, this technique has never been implemented and used in a multilevel framework. Hence, we provide a modified version of the wild bootstrap that could be useful in cases of hierarchical and heteroscedastic data. The next subsection is devoted to this matter.

\subsection{The wild bootstrap for multilevel models}\label{subsec:wild_multi}
Now, we adapt the procedure presented above to the case of hierarchical data. Consider the classic multilevel model in Eq. (\ref{eq:general model}) but for the ($n_j\times 1$) response of the generic group $j$:
\begin{equation}\label{eq:mod_vector}
\y_j=\X_j\boldsymbol{\beta} + \boldsymbol{\nu}_j, \quad \text{with} \quad \boldsymbol{\nu}_j=\Z_j\mathbf{u}_j+\boldsymbol{\epsilon}_j,
\end{equation}
for all $j=1,\ldots,J$. Note that the wild bootstrap procedure requires the disturbances to be mutually independent. However, in the multilevel framework the compound error terms of two generic units in the group $j$, $\nu_{ij}=z_{ij}^{\T}\mathbf{u}_j+\epsilon_{ij}$ and $\nu_{i'j}=z_{i'j}^{\T}\mathbf{u}_j+\epsilon_{i'j}$, are not independent by definition. Instead, the error terms corresponding to the whole generic group $j$, $\boldsymbol{\nu}_j$,  are mutually independent. Therefore, handling the vectorial form of the multilevel model (\ref{eq:mod_vector}), rather than the univariate form (\ref{eq:mod}), allows from one hand to put oneself in the same situation of the classic wild bootstrap procedure, and, from another hand, to take into account the intra-class correlation of hierarchical data.\\
Denoting with $\mathbf{H}_j=\X_j(\X^{\T}\X)^{-1}\X_j^{T}$ the $j$-th diagonal block of the orthogonal projection matrix associated to the design matrix $\mathbf{H}$, the HCCME expressions in (\ref{eq:HC}) become
\begin{align}\label{eq:HC2}
\text{HC}_2:\quad \tilde{\mathbf{v}}_j&=\text{diag}\big(\mathbf{I}_{n_j}-\mathbf{H}_j\big)^{-1/2} \circ \hat{\mathbf{v}}_j \nonumber \\
\text{HC}_3: \quad \tilde{\mathbf{v}}_j&=\text{diag}\big(\mathbf{I}_{n_j}-\mathbf{H}_j\big) \circ \hat{\mathbf{v}}_j,\end{align}
where the vector of the residuals for the $j$-th group is computed as \[\hat{\mathbf{v}}_j= \y_j - \X_j\boldsymbol{\hat{\beta}}\] and the operator ``$\circ$'' denotes the Hadamard (or entrywise) product.
\par
We suggest to implement the wild bootstrap for a multilevel model as follows:
 \begin{enumerate}
 \item draw $j$ independent values, $w_j$, for $i=j,\ldots,J$, from an auxiliary distribution with zero mean and unit variance such as $F_1$ and $F_2$;
 \item generate the bootstrap samples
 \[\y_j^*=\X_j\boldsymbol{\hat{\beta}} + \mathbf{\tilde{v}}_jw_j,\] where the transformed residuals $\mathbf{\tilde{v}}_j$ are as in (\ref{eq:HC2});
  \item compute the bootstrap value $\boldsymbol{\hat{\theta}}^*$ on the generated sample;
  \item repeat steps $1-3$ B times as to obtain B bootstrap replications of the parameters.
 \end{enumerate}

\section{Monte Carlo study}\label{sec:MC}
In this section we present the results of a Monte Carlo study that compares the three bootstrap procedures described above (Section \ref{sec:boot}) with the adapted wild bootstrap procedure for multilevel data (Section \ref{subsec:wild_multi}). Here, the wild bootstrap is implemented with $F_1$ as in Eq.~(\ref{eq:F1}) as auxiliary distribution and HC$_2$ as in Eq.~(\ref{eq:HC2}) as HCCME form. Further, we study the behaviour of the wild bootstrap with different choices of auxiliary distributions and HCCME forms. Consider the following two-level model
\begin{equation}
\label{eq:mod}y_{ij}=\beta_0  + u_{0j} + (\beta_1+u_{1j})x_{1ij} +  \epsilon_{ij},
\end{equation}
for level-1 units $i=1,\ldots,n_j$ and level-2 units $j=1,\ldots,J$. In this study, the clusters contain the same number of level-2 units, that is $n_j=n$ for all $j$. The true values chosen for the regression coefficients are $\beta_0=3$ and $\beta_1=5$ and the $x_{1ij}$'s are simulated from a standard normal distribution.
\par
In order to assess the finite size performance of the bootstrap schemes in presence of non-constant variance we generate samples with both homoscedastic and heteroscedastic level-1 errors, $\epsilon_{ij}$. To this aim, we define
\[
\epsilon_{ij}=s_{ij}\nu_{ij} \quad \forall i,j
\]
and set $s_{ij}=1$ for all $i,j$ for generating homoscedastic data, and $s_{ij}=x_{1ij}$ for obtaining heteroscedastic data. Hence, the variance of level-1 error terms is $s_{ij}^2\sigma^2_{\nu}$, where $\sigma^2_{\nu}=\mbox{V}(\nu_{ij})$ for all $i$ and $j$. Also, we control for deviations from normality by adopting two different error distributions. In the first case, we draw $N$ values of $\nu_{ij} \sim N(0,\sigma^2_\nu=2)$ and, for each $j=1,\ldots,J$, we draw a sample $(u_{0j},u_{1j})$ from the bivariate normal distribution: 
\[
\left[\begin{array}{c} u_{0j} \\ u_{1j} \end{array}\right] \sim N\Bigg(\left[\begin{array}{c} 0 \\ 0 \end{array}\right],
\left[\begin{array}{cc} \sigma^2_{u0}=2 & \sigma_{u01}=0.5\\ \sigma_{u01}=0.5 & \sigma^2_{u1}=2\\
                                                 \end{array}\right]\Bigg).
\]
As for the non Gaussian case, we draw $N$ values of $\nu_{ij}$ a from $\chi_1^2 -1$ distribution and, for each $j=1,\ldots,J$, we draw a sample $(h_1,h_2)$ from the bivariate normal distribution
 \[
 \left[\begin{array}{c} h_{1} \\ h_{2} \end{array}\right] \sim
 N\Bigg(\left[\begin{array}{c} 0 \\ 0
                      \end{array}\right], \left[\begin{array}{cc} 1 & 0.5\\ 0.5 & 1\\
                                                 \end{array}\right]\Bigg);
                                                 \]
then, we set $u_{0j}=h_1^2-1$ and $u_{1j}=h_2^2-1$ so that both random effects have marginal $(\chi_1^2-1)$-distribution with mean 0, variances $\sigma^2_{u0}$ and $\sigma^2_{u1}$ equal to 2 and covariance $\sigma_{u01}$ equal to 0.5.
\par
In the study, we vary also the group size $n$ and the number of groups $J$ as follows:
\[
\begin{array}{|r|c|c|c|}
\hline
  & n &  J & N  \\
\hline
 1 & 10 & 20 & 200 \\
 2 & 10 & 80 & 800 \\
 3 & 20 & 40 & 800 \\
 4 & 40 & 80 & 3200 \\
\hline
\end{array}
\]
\par
Note that settings 1 and 2 differ in the number of groups whereas settings 2 and 4 differ in the group size. For each setting, we simulate 500 datasets from model (\ref{eq:mod}), and for each dataset we compute restricted maximum likelihood estimates of the parameters. The number of bootstrap replications is $B=999$, hence we obtain $B$ replications of the parameters, $\boldsymbol{\hat{\theta}}^*=\{\hat{\beta}_0^*, \hat{\beta}_1^*,(\hat{\sigma}^2_{\epsilon})^*,(\hat{\sigma}^2_{u0})^*,(\hat{\sigma}_{u01})^*,(\hat{\sigma}^2_{u1})^*\}$. Finally, we derive 95\% bootstrap confidence intervals \citep{Davison,Leeden} by sorting the bootstrap replications and taking the following two percentiles
 \[\bigg[\hat{\theta}^*_{100\,\alpha/2},\hat{\theta}^*_{100\,(1-\alpha/2)}\bigg]\]
 with $\alpha=0.05$. As measures of performance we take the empirical coverage of the confidence interval and its average length. Table~\ref{tab:scheme} summarizes the scenarios of the simulation study indicating the tables of the results for each case.

\begin{table}[ht]
\caption{Summary of the scenarios of the Monte Carlo study.}\label{tab:scheme}
\centering
\medskip
\begin{tabular}{|l|l|l|}
\hline
\emph{Errors}& Homoscedastic &  Heteroscedastic  \\
\hline
 Gaussian & Scenario 1: Table \ref{tab:norm_homo}& Scenario 2: Table \ref{tab:norm_het}\\
$\chi_1^2-1$& Scenario 3: Table \ref{tab:chisq_homo} & Scenario 4: Table \ref{tab:chisq_het}\\
\hline
\end{tabular}
\end{table}

In the case of Gaussian and homoscedastic errors (scenario 1, Table \ref{tab:norm_homo}), all the bootstrap procedures considered behave quite well.  In particular, for regression coefficients, the residual bootstrap has the best performance (even though by a tiny margin) in all the four sample sizes. As for the estimation of $\sigma^2_{\epsilon}$, the wild bootstrap always produces a coverage of 100\% at the price of wider confidence intervals. Moreover, for level-2 parameters, the intervals based on the parametric bootstrap have the highest coverage when both the level-1 and level-2 sample sizes are low. However, as the sample sizes increases, the wild bootstrap behaves better other schemes for almost all parameters. When the level-1 errors are Gaussian but heteroscedastic (scenario 2, Table \ref{tab:norm_het}), the coverage of the parametric bootstrap is good for all but the within-group variance  $\sigma^2_{\epsilon}$. The same happens for the residual and the wild bootstrap but with some important differences. In fact, the wild bootstrap has always the highest coverage when the sample sizes increase. Also, as in scenario 1, the coverage for the within-group variance $\sigma^2_{\epsilon}$ is rather low for all the schemes but the wild bootstrap. In the case of homoscedastic errors with $\chi_1^2-1$ distribution (scenario 3, Table \ref{tab:chisq_homo}), the parametric bootstrap produces a small coverage not only for the within-group variance but also for level-2 parameters. On the other hand, the coverage of the residual bootstrap is still good, in line with the results in \citet{Carpenter}. However, when the sample sizes increase, the wild bootstrap produces again the highest coverage in every instance. Lastly, when the $\chi_1^2-1$-errors are heteroscedastic (scenario 4, Table \ref{tab:chisq_het}), the results are not clear cut as there is not a definite winner. The parametric bootstrap has low coverage mainly when variances are involved. The residual bootstrap has a fair performance, but, again, with big sample sizes the wild bootstrap assures good coverage and length for all the parameters.  Finally, note that the cases bootstrap is always outperformed by other procedures, as happens in linear regression models (see among the others \citet{Davison} and \citet{Flachaire99}).
\par
\begin{figure}
 \psfrag{sigma2_eps}{\small $\sigma^2_{\epsilon}$}
  \centering
  \includegraphics[width=7cm,height=7cm,keepaspectratio=true]{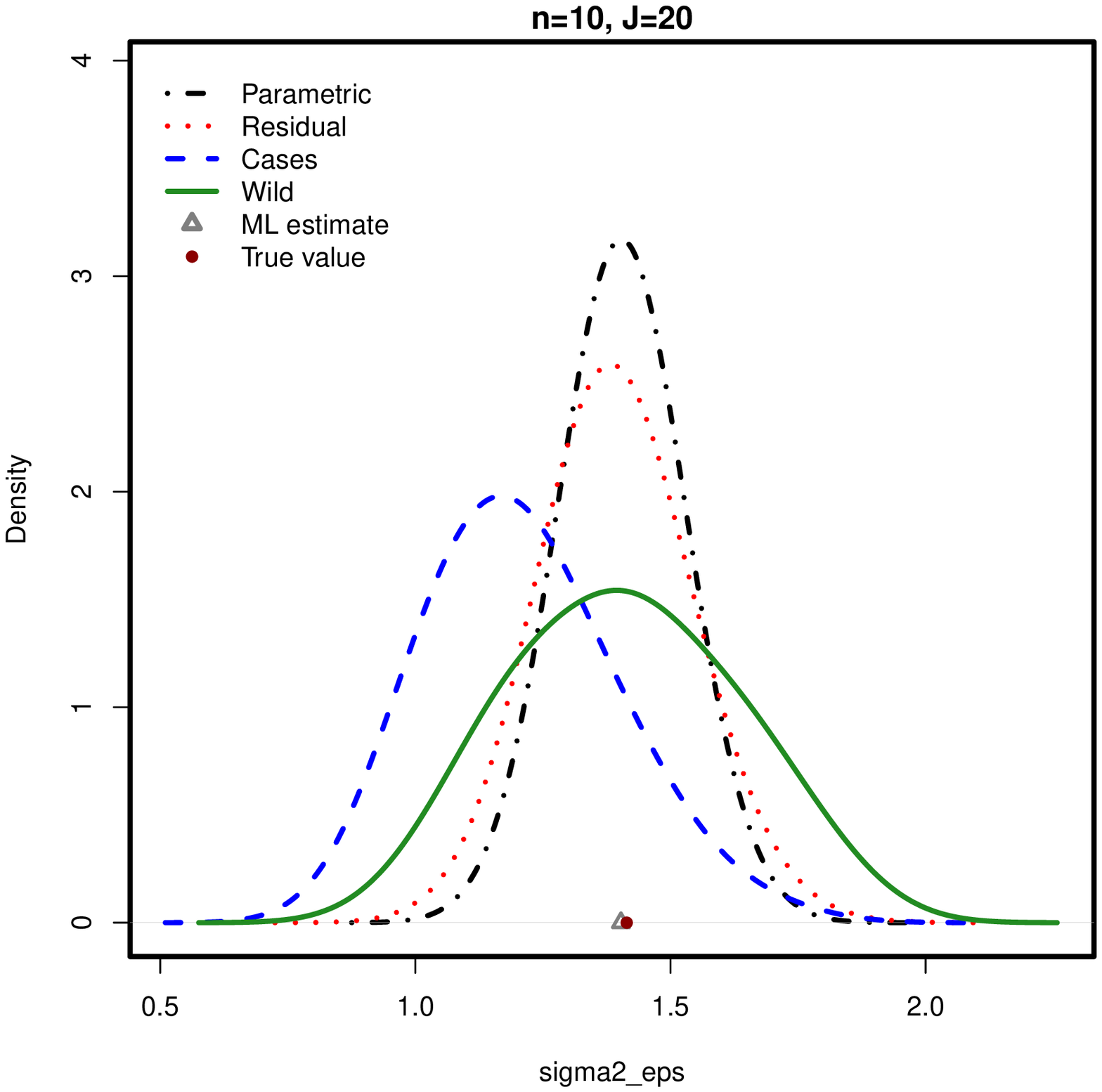}
  \includegraphics[width=7cm,height=7cm,keepaspectratio=true]{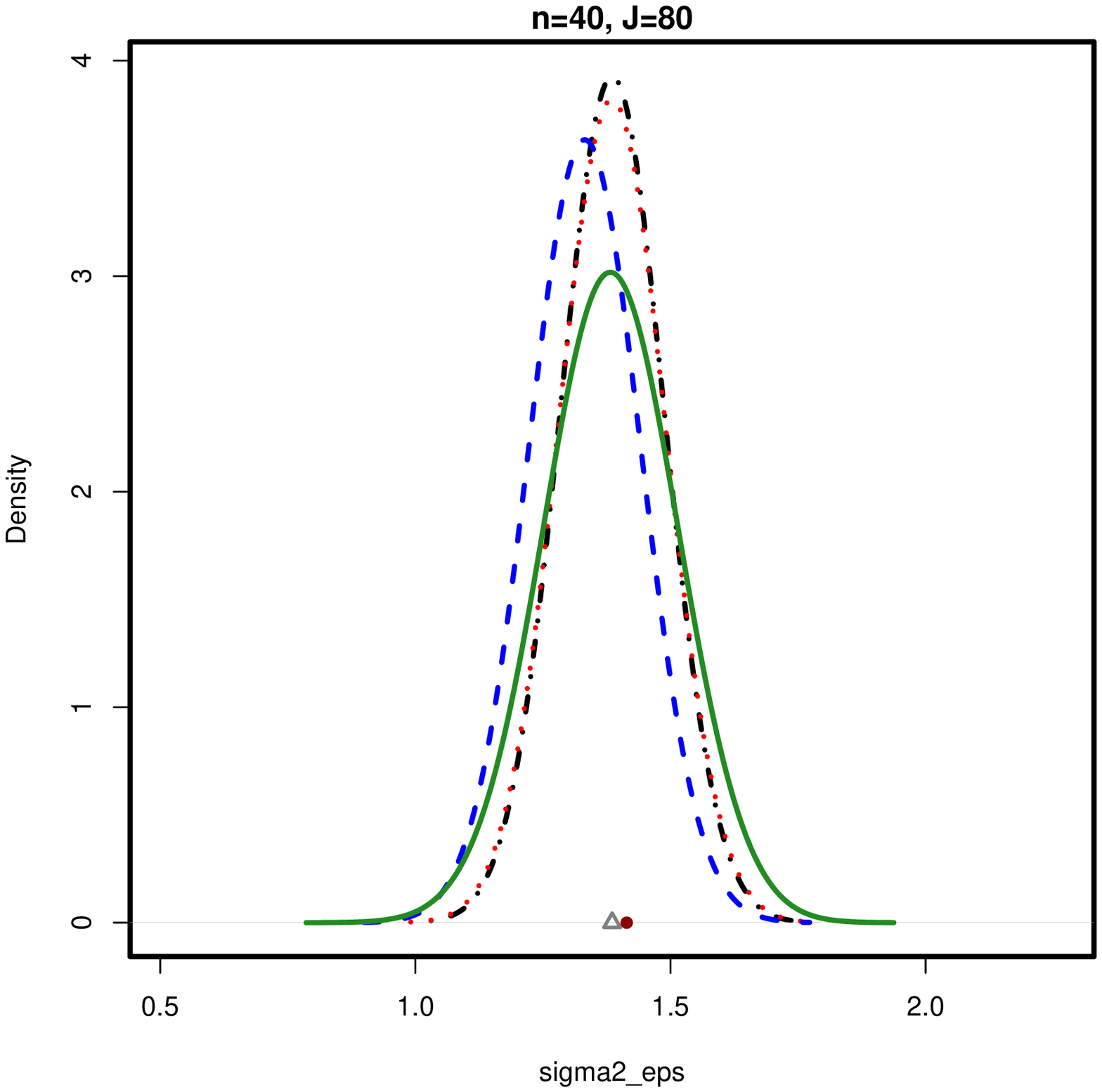}
    \caption{\label{fig:sigma} Distributions of the bootstrap replications for the within-group variance $\sigma^2_{\epsilon}$ for two samples with heteroscedastic Gaussian errors: (left) small sample size ($n=10$, $J=20)$; (right) big sample size ($n=80$, $J=80)$.}
\end{figure}
\begin{figure}
\centering
 \psfrag{sigma_u01}{\small $\sigma_{u01}$}
  \includegraphics[width=7cm,height=7cm,keepaspectratio=true]{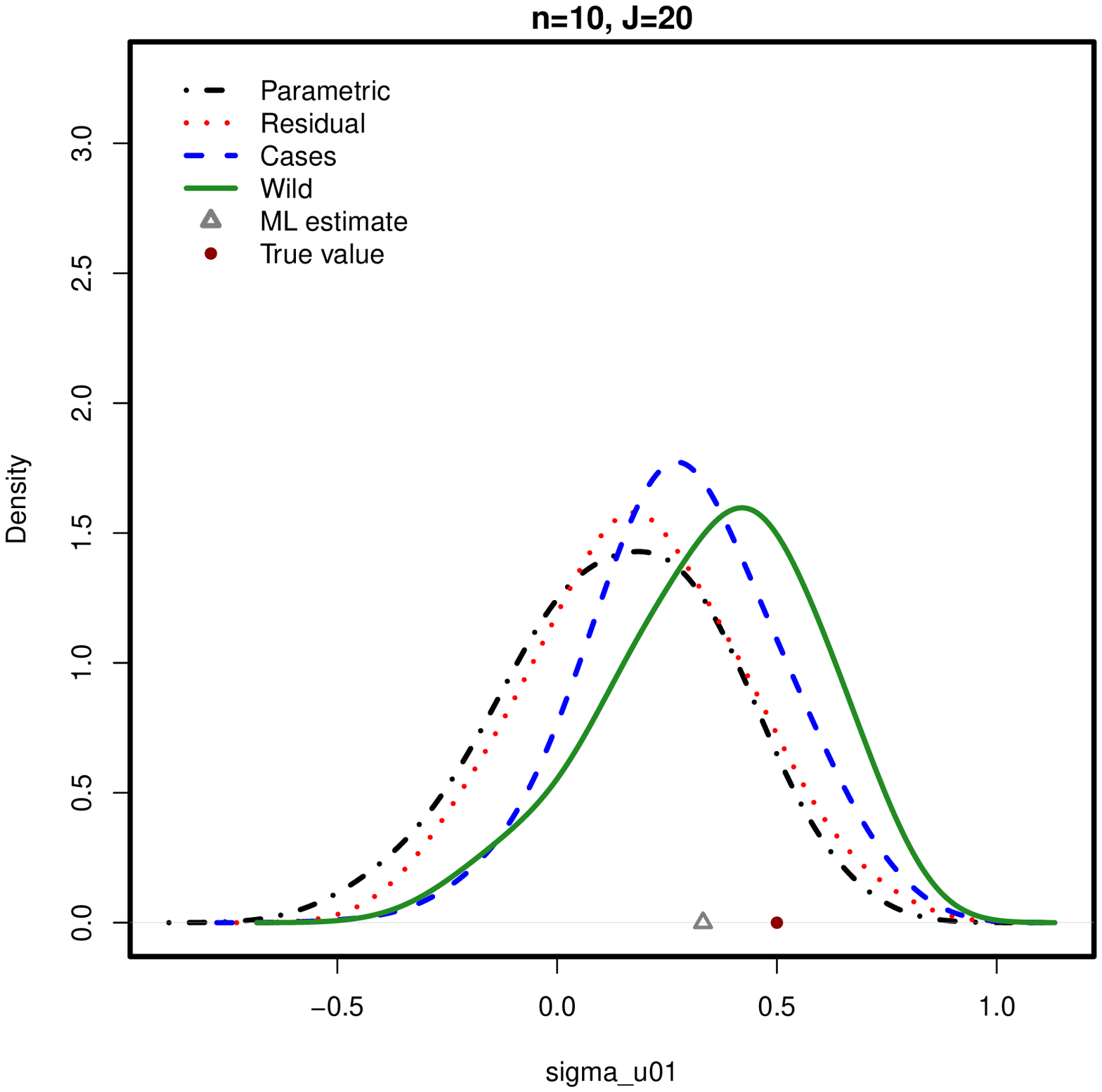}
  \includegraphics[width=7cm,height=7cm,keepaspectratio=true]{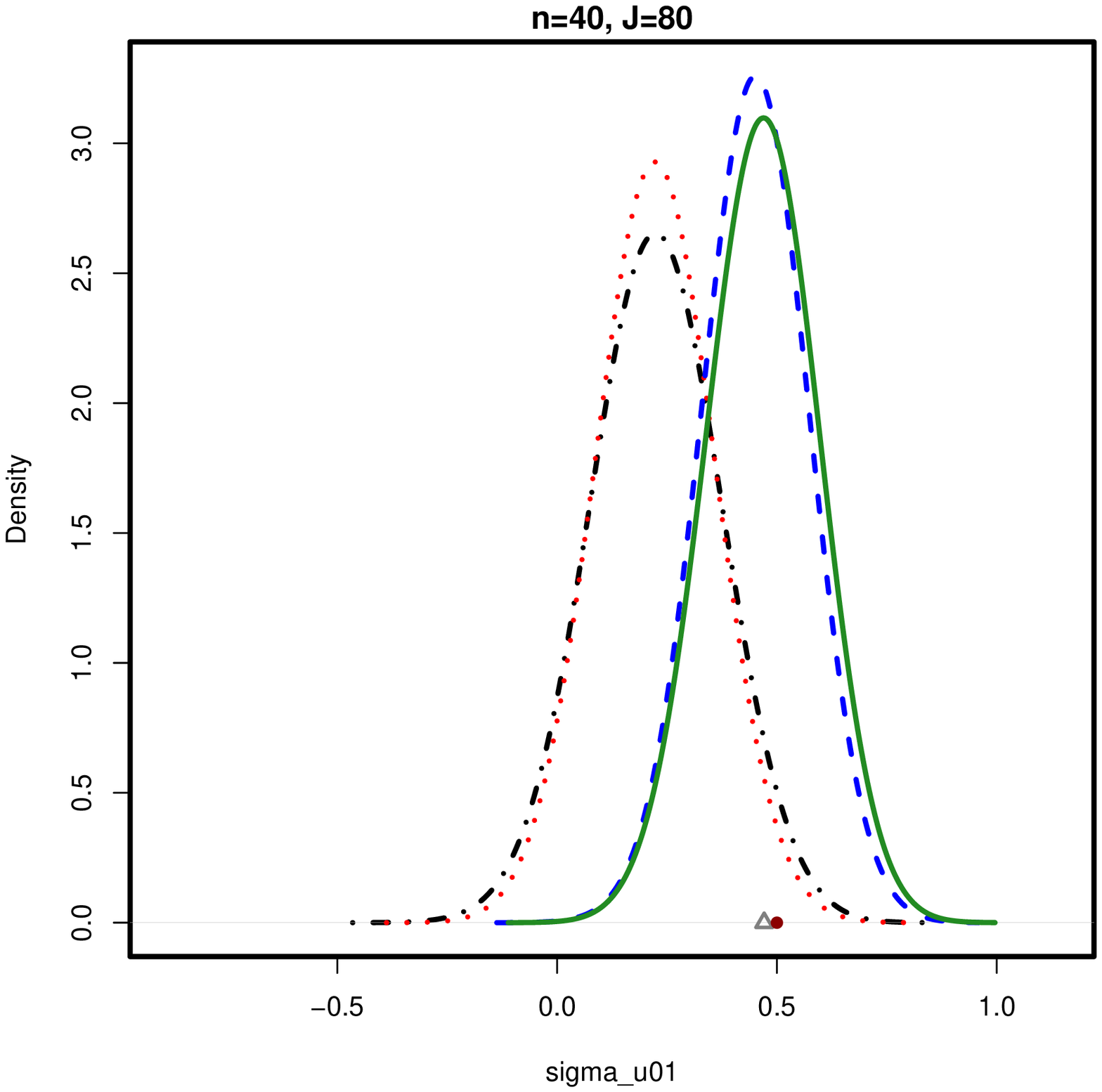}
\caption{\label{fig:sigma_u01}Distributions of the bootstrap replications for the covariance between the random slope and the random intercept $\sigma_{u01}$ for two samples with heteroscedastic Gaussian errors: (left) small sample size ($n=10$, $J=20)$; (right) big sample size ($n=80$, $J=80)$.}
\end{figure}

In Figures \ref{fig:sigma} and \ref{fig:sigma_u01} we provide a visual comparison of the bootstrap methods in the heteroscedastic scenario.  The figures show a single instance of the distributions of the bootstrap replications for $\sigma^2_{\epsilon}$ and $\sigma_{u01}$, respectively where the samples have heteroscedastic Gaussian errors. The left panels show the result for small sample size ($N =200$) whereas the right panels refer to a big sample size ($N = 3200$). Figure \ref{fig:sigma} shows that, in this example, both the parametric and the residual bootstrap behave well for $\sigma^2_{\epsilon}$.  On the contrary, as shown in Figure \ref{fig:sigma_u01}, for the covariance $\sigma_{u01}$ the wild bootstrap outperforms the other methods, both for small and bug sample sizes.
\par
Now, we focus on the impact of different choices of HCCME forms and auxiliary distributions on the performance of the wild bootstrap. Table \ref{tab:wild} reports the results of a simulation study that investigates the coverage of bootstrap percentile intervals for different versions of the wild bootstrap in the same four scenarios as before. The sample sizes considered here are $n=20$ and $J=40$. The results show clearly that the best behaved version of the wild bootstrap in all the scenarios is the one with the auxiliary distribution $F_1$ (Eq.~\ref{eq:F1}) and HC$_3$ (Eq.~\ref{eq:HC2}) as HCCME.
These results are somehow in contrasts with those of \citet{Davidson} which show that for a classic regression model the distribution $F_2$ is always the best choice.

\begin{table}
\scriptsize
\caption{Scenario 1: Gaussian and homoscedastic lev-1 errors. Coverage (\%) and average length (in parenthesis) for $95\%$ bootstrap confidence intervals.}\label{tab:norm_homo}
\centering
\medskip
\begin{tabular}{|c|cccc|cccc|}
\hline
             & \textbf{Parametric} & \textbf{Residual} & \textbf{Cases} & \textbf{Wild} & \textbf{Parametric} & \textbf{Residual} & \textbf{Cases} & \textbf{Wild}\\
\hline
&\multicolumn{4}{|c}{$n=10,\quad J=20, \quad N=200$} & \multicolumn{4}{|c|}{$n=20,\quad J=40, \quad N=800$}\\
\hline
 $\beta_0$                                                               &  95.2 &  95.2 & 90.6 & 94.2                     & 94 & 94 & 85 & 93 \\
                                                                         & (1.32) & (1.32) & (1.1) & (1.29)                & (0.89) & (0.89) & (0.71) & (0.88) \\
 $\beta_1$                                                               & 94.4 & 94.8 & 90.4 & 93.6                       & 93.6 & 94.4 & 87.4 & 93.4 \\
                                                                         & (1.33) & (1.32) & (1.14) & (1.3)                & (0.89) & (0.89) & (0.72) & (0.88) \\
 $\sigma^2_{\epsilon}$                                                   & 94.4 & 94.2 & 72.2 & 100                        & 94.6 & 94 & 72.6 & 100 \\
                                                                         & (0.31) & (0.31) & (0.39) & (0.7)                & (0.15) & (0.15) & (0.19) & (0.46) \\
  $\sigma^2_{u0}$                                                        & 94.2 & 92 & 94.2 & 93.6                         & 92.8 & 91.2 & 90 & 94.4 \\
                                                                         & (1.01) & (0.96) & (0.88) & (1)                  & (0.65) & (0.63) & (0.53) & (0.73) \\
 $\sigma_{u01}$                                                          & 97.4 & 96.6 & 92.4 & 89.6                       & 96.2 & 95 & 88.6 & 92.8 \\
                                                                          & (2.09) & (1.97) & (1.96) & (1.97)              & (1.32) & (1.27) & (1.12) & (1.3) \\
 $\sigma^2_{u1}$                                                          & 91.6 & 90.4 & 93.2 & 91.4                      & 92.2 & 91 & 89 & 93.6 \\
                                                                          & (1.03) & (0.99) & (0.95) & (1.02)               & (0.66) & (0.63) & (0.54) & (0.73) \\
\hline
& \multicolumn{4}{|c}{$n=10,\quad J=80, \quad N=800$} & \multicolumn{4}{|c|}{$n=40,\quad J=80, \quad N=3200$}\\
\hline
  $\beta_0$                                                              & 94.8 & 94.8 & 90.2 & 94.6                     & 94.4 & 94.6 & 87.6 & 93.8 \\
                                                                         & (0.65) & (0.66) & (0.55) & (0.65)             & (0.63) & (0.63) & (0.49) & (0.62) \\
  $\beta_1$                                                              & 93.2 & 93.6 & 89.6 & 93.4                     & 95 &95.4 & 88.6 & 95 \\
                                                                         & (0.66) & (0.66) & (0.56) & (0.65)             & (0.63) & (0.63) & (0.49) & (0.62) \\
  $\sigma^2_{\epsilon}$                                                  & 95 & 95 & 13 & 100                            & 94.8 & 94.4 & 75.4 & 100 \\
                                                                         & (0.16) & (0.16) & (0.2) & (0.35)              & (0.07) & (0.07) & (0.1) & (0.32) \\
   $\sigma^2_{u0}$                                                       & 93 & 92 & 90.6 & 95.6                         & 93.6 & 92.6 & 88 & 95.8 \\
                                                                         & (0.49) & (0.48) & (0.42) & (0.56)             & (0.45) & (0.44) & (0.35) & (0.53) \\
  $\sigma_{u01}$                                                         & 95.4 & 94.2 & 93.2 & 94.2                     & 94.2 & 91.2 & 87.8 & 94.8 \\
                                                                         & (0.99) & (0.98) & (0.94) & (1.01)             & (0.91) & (0.89) & (0.74) & (0.93) \\
  $\sigma^2_{u1}$                                                        & 92.6 & 92 & 90.6 & 94.8                       & 94 & 94 & 88.8 & 97.6 \\
                                                                         & (0.5) & (0.49) & (0.45) & (0.57)              & (0.45) & (0.44) & (0.36) & (0.53) \\
   \hline
\end{tabular}
\end{table}

\begin{table}
\scriptsize
\caption{Scenario 2: Gaussian and heteroscedastic lev-1 errors. Coverage (\%) and average length (in parenthesis) for $95\%$ bootstrap confidence intervals.}\label{tab:norm_het}
\centering
\medskip
\begin{tabular}{|c|cccc|cccc|}
\hline
             & \textbf{Parametric} & \textbf{Residual} & \textbf{Cases} & \textbf{Wild} & \textbf{Parametric} & \textbf{Residual} & \textbf{Cases} & \textbf{Wild}\\
\hline
&\multicolumn{4}{|c}{$n=10,\quad J=20, \quad N=200$} & \multicolumn{4}{|c|}{$n=20,\quad J=40, \quad N=800$}\\
\hline
 $\beta_0$                                                                & 94.8 & 94.6 & 88.2 & 93.8 & 92.2 & 92.2 & 83.8 & 91.2 \\
                                                                          & (1.32) & (1.32) & (1.1) & (1.27) & (0.89) & (0.89) & (0.71) & (0.87) \\
 $\beta_1$                                                                & 94.4 & 94.2 & 91.2 & 93.8 & 94.2 & 94.2 & 90.2 & 93.8 \\
                                                                          & (1.33) & (1.32) & (1.14) & (1.36) & (0.89) & (0.89) & (0.72) & (0.92) \\
 $\sigma^2_{\epsilon}$                                                    & 48.2 & 65.8 & 47.6 & 89.4 & 50 & 69 & 49.2 & 98.6 \\
                                                                          & (0.31) & (0.31) & (0.39) & (0.73) & (0.15) & (0.15) & (0.19) & (0.49) \\
  $\sigma^2_{u0}$                                                         & 92.4 & 91.6 & 90.8 & 92.6 & 91.4 & 88.4 & 85.4 & 93 \\
                                                                          & (1.01) & (0.96) & (0.88) & (0.99) & (0.65) & (0.63) & (0.53) & (0.72) \\
 $\sigma_{u01}$                                                           & 98 & 97.2 & 93.2 & 91.4 & 96.4 & 95.8 & 90.8 & 92.6 \\
                                                                           & (2.09) & (1.97) & (1.96) & (2.05) & (1.32) & (1.27) & (1.12) & (1.34) \\
 $\sigma^2_{u1}$                                                           & 95.6 & 93 & 94.6 & 95.4 & 95 & 92.4 & 90.6 & 96.4 \\
                                                                           & (1.03) & (0.99) & (0.95) & (1.06) & (0.66) & (0.63) & (0.54) & (0.76) \\
\hline
& \multicolumn{4}{|c}{$n=10,\quad J=80, \quad N=800$} & \multicolumn{4}{|c|}{$n=40,\quad J=80, \quad N=3200$}\\
\hline
  $\beta_0$                                                               & 95.4 & 95.6 & 88.8 & 95.8 & 93.8 & 94 & 86.6 & 94 \\
                                                                          & (0.65) & (0.66) & (0.55) & (0.64) & (0.63) & (0.63) & (0.49) & (0.62) \\
  $\beta_1$                                                               & 95.8 & 96 & 92.2 & 95.2 & 95.4 & 95 & 88.4 & 94.8 \\
                                                                          & (0.66) & (0.66) & (0.56) & (0.69) & (0.63) & (0.63) & (0.49) & (0.63) \\
  $\sigma^2_{\epsilon}$                                                   & 20.6 & 37.6 & 5.4 & 77.4 & 45.2 & 72.6 & 50.8 & 100 \\
                                                                          & (0.16) & (0.16) & (0.2) & (0.38) & (0.07) & (0.07) & (0.1) & (0.33) \\
   $\sigma^2_{u0}$                                                        & 94.6 & 93.2 & 92.4 & 96.8 & 94 & 93 & 85.6 & 96.6 \\
                                                                          & (0.49) & (0.48) & (0.42) & (0.56) & (0.45) & (0.44) & (0.35) & (0.53) \\
  $\sigma_{u01}$                                                          & 95.2 & 95.6 & 93.6 & 94.8 & 93.8 & 91.4 & 90.2 & 95.6 \\
                                                                          & (0.99) & (0.98) & (0.94) & (1.05) & (0.91) & (0.89) & (0.74) & (0.95) \\
  $\sigma^2_{u1}$                                                         & 90.2 & 89.4 & 73.8 & 96.2 & 97 & 96.2 & 91.2 & 99 \\
                                                                          & (0.5) & (0.49) & (0.45) & (0.59) & (0.45) & (0.44) & (0.36) & (0.54) \\
   \hline
\end{tabular}
\end{table}

\begin{table}
\scriptsize
\caption{Scenario 3: $\chi_1^2-1$ and homoscedastic lev-1 errors. Coverage (\%) and average length (in parenthesis) for $95\%$ bootstrap confidence intervals.}\label{tab:chisq_homo}
\centering
\medskip
\begin{tabular}{|c|cccc|cccc|}
\hline
             & \textbf{Parametric} & \textbf{Residual} & \textbf{Cases} & \textbf{Wild} & \textbf{Parametric} & \textbf{Residual} & \textbf{Cases} & \textbf{Wild}\\
\hline
&\multicolumn{4}{|c}{$n=10,\quad J=20, \quad N=200$} & \multicolumn{4}{|c|}{$n=20,\quad J=40, \quad N=800$}\\
\hline
 $\beta_0$                                                               & 89.4 & 89.8 & 86.6 & 89.6                          & 92.2 & 92.4 & 85 & 91.6 \\
                                                                         & (1.26) & (1.25) & (1.04) & (1.19)                  & (0.88) & (0.87) & (0.69) & (0.85) \\
 $\beta_1$                                                               & 89.6 & 89.8 & 88.6 & 90                            & 91 & 91.6 & 84.6 & 91.8 \\
                                                                         & (1.26) & (1.25) & (1.08) & (1.19)                  & (0.87) & (0.87) & (0.69) & (0.84) \\
 $\sigma^2_{\epsilon}$                                                   & 61.2 & 84.6 & 81.4 & 96.8                           & 54.8 & 91.4 & 90.6 & 99.6 \\
                                                                         & (0.31) & (0.57) & (0.77) & (0.85)                  & (0.15) & (0.33) & (0.46) & (0.56) \\
  $\sigma^2_{u0}$                                                        & 60.6 & 70.8 & 74.8 & 65.2                          & 56.4 & 75.2 & 69 & 73 \\
                                                                         & (0.98) & (1.33) & (1.21) & (1.11)                     & (0.65) & (1.08) & (0.88) & (0.95) \\
 $\sigma_{u01}$                                                          & 84.2 & 82 & 80.6 & 66.4                              & 85.4 & 81.8 & 78 & 78 \\
                                                                          & (1.94) & (2.02) & (2.19) & (2.01)                  & (1.27) & (1.49) & (1.28) & (1.44) \\
 $\sigma^2_{u1}$                                                          & 61.8 & 68.2 & 79.6 & 66.2                          & 59.6 & 74.8 & 71.6 & 74.2 \\
                                                                          & (0.99) & (1.25) & (1.3) & (1.12)                   & (0.64) & (1.01) & (0.87) & (0.93) \\
\hline
& \multicolumn{4}{|c}{$n=10,\quad J=80, \quad N=800$} & \multicolumn{4}{|c|}{$n=40,\quad J=80, \quad N=3200$}\\
\hline
  $\beta_0$                                                               & 94.6 & 95.2 & 91.4 & 95.2                           & 93.2 & 94.2 & 84.4 & 93 \\
                                                                          & (0.64) & (0.64) & (0.53) & (0.63)                   & (0.61) & (0.61) & (0.47) & (0.6) \\
  $\beta_1$                                                               & 94.2 & 95.2 & 89.4 & 94.6                            & 92.2 & 93.6 & 85 & 93.8 \\
                                                                          & (0.65) & (0.65) & (0.55) & (0.64)                      & (0.61) & (0.61) & (0.47) & (0.6) \\
  $\sigma^2_{\epsilon}$                                                   & 57.4 & 89.6 & 71 & 98.4                               & 55.8 & 94.4 & 95.8 & 100 \\
                                                                          & (0.15) & (0.32) & (0.44) & (0.47)                     & (0.07) & (0.18) & (0.25) & (0.36) \\
   $\sigma^2_{u0}$                                                        & 62.8 & 84 & 87 & 84.6                                   & 55.2 & 83 & 73.6 & 83.4 \\
                                                                          & (0.48) & (0.85) & (0.72) & (0.84)                      & (0.44) & (0.88) & (0.67) & (0.82) \\
  $\sigma_{u01}$                                                          & 7.6 & 82.4 & 85.6 & 54.2                                & 76 & 76.8 & 72.4 & 77.2 \\
                                                                          & (0.5) & (1.19) & (1.16) & (0.51)                          & (0.87) & (1.13) & (0.91) & (1.12) \\
  $\sigma^2_{u1}$                                                         & 59.4 & 79 & 87.8 & 82.6                                 & 54.4 & 82.2 & 71 & 82.8 \\
                                                                          & (0.49) & (0.83) & (0.76) & (0.84)                      & (0.44) & (0.83) & (0.66) & (0.8) \\
   \hline
\end{tabular}
\end{table}
\begin{table}
\scriptsize
\caption{Scenario 4: $\chi_1^2-1$ and heteroscedastic lev-1 errors. Coverage (\%) and average length (in parenthesis) for $95\%$ bootstrap confidence intervals.}\label{tab:chisq_het}
\centering
\medskip
\begin{tabular}{|c|cccc|cccc|}
\hline
             & \textbf{Parametric} & \textbf{Residual} & \textbf{Cases} & \textbf{Wild} & \textbf{Parametric} & \textbf{Residual} & \textbf{Cases} & \textbf{Wild}\\
\hline
&\multicolumn{4}{|c}{$n=10,\quad J=20, \quad N=200$} & \multicolumn{4}{|c|}{$n=20,\quad J=40, \quad N=800$}\\
\hline
 $\beta_0$                                                            & 90 & 89.6 & 84.8 & 89                                  & 93 & 93 & 84 & 92.6 \\
                                                                      & (1.26) & (1.25) & (1.04) & (1.19)                      & (0.88) & (0.87) & (0.69) & (0.84) \\
 $\beta_1$                                                            & 90.6 & 91.8 & 89.2 & 90.6                              & 91.6 & 91.2 & 87 & 91.4 \\
                                                                      & (1.26) & (1.25) & (1.08) & (1.27)                      & (0.87) & (0.87) & (0.69) & (0.88) \\
 $\sigma^2_{\epsilon}$                                                & 33.8 & 62.6 & 60.8 & 78.4                                & 30.6 & 72.8 & 73.2 & 90 \\
                                                                      & (0.31) & (0.57) & (0.77) & (0.87)                       & (0.15) & (0.33) & (0.46) & (0.63) \\
  $\sigma^2_{u0}$                                                     & 63.2 & 70.4 & 74 & 65.6                                & 54.6 & 76.2 & 68.8 & 74.6 \\
                                                                      & (0.98) & (1.33) & (1.21) & (1.1)                        & (0.65) & (1.08) & (0.88) & (0.95) \\
 $\sigma_{u01}$                                                       & 84.4 & 80.8 & 81.2 & 66.8                                & 88 & 83.4 & 81.2 & 78.6 \\
                                                                       & (1.94) & (2.02) & (2.19) & (2.09)                        & (1.27) & (1.49) & (1.28) & (1.48) \\
 $\sigma^2_{u1}$                                                       & 69 & 76.2 & 86.6 & 72.8                                & 65.4 & 78.4 & 81.4 & 78.4 \\
                                                                       & (0.99) & (1.25) & (1.3) & (1.13)                         & (0.64) & (1.01) & (0.87) & (0.94) \\
\hline
& \multicolumn{4}{|c}{$n=10,\quad J=80, \quad N=800$} & \multicolumn{4}{|c|}{$n=40,\quad J=80, \quad N=3200$}\\
\hline
  $\beta_0$                                                               & 93 & 93.4 & 89.6 & 92.6                        & 92.8 & 93 & 83.6 & 93 \\
                                                                          & (0.64) & (0.64) & (0.53) & (0.62)              & (0.61) & (0.61) & (0.47) & (0.6) \\
  $\beta_1$                                                               & 94.8 & 94.2 & 91.4 & 94.4                       & 93.6 & 93.4 & 87.4 & 93.2 \\
                                                                          & (0.65) & (0.65) & (0.55) & (0.67)              & (0.61) & (0.61) & (0.47) & (0.61) \\
  $\sigma^2_{\epsilon}$                                                   & 18.6 & 56.2 & 40.8 & 74.8                          & 27.6 & 84 & 84 & 95.4 \\
                                                                          & (0.15) & (0.32) & (0.44) & (0.53)                 & (0.07) & (0.18) & (0.25) & (0.41) \\
   $\sigma^2_{u0}$                                                        & 62.2 & 85 & 83.2 & 84.8                            & 56.6 & 83.8 & 73.2 & 82.4 \\
                                                                          & (0.48) & (0.85) & (0.72) & (0.84)                  & (0.44) & (0.88) & (0.67) & (0.82) \\
  $\sigma_{u01}$                                                          & 2.8 & 83 & 85.2 & 44.6                              & 76.6 & 77 & 75.8 & 79 \\
                                                                          & (0.5) & (1.19) & (1.16) & (0.5)                        & (0.87) & (1.13) & (0.91) & (1.14) \\
  $\sigma^2_{u1}$                                                         & 69.2 & 90.4 & 91.6 & 89.8                               & 60.4 & 85 & 79.2 & 84.8 \\
                                                                          & (0.49) & (0.83) & (0.76) & (0.83)                       & (0.44) & (0.83) & (0.66) & (0.8) \\
   \hline
\end{tabular}
\end{table}
\begin{table}[!htbp]\scriptsize
\caption[Scenarios 1 and 2 of the MC study]{Simulation study with different versions of the wild bootstrap ($n=20$ and $J=40$).}\label{tab:wild}
\centering
\medskip
\begin{tabular}{|c|c|cccc|cccc|}
\hline

\multicolumn{2}{|c}{}&\multicolumn{2}{|c}{HC$_2$} & \multicolumn{2}{c|}{HC$_3$}& \multicolumn{2}{|c}{HC$_2$} & \multicolumn{2}{c|}{HC$_3$}\\
\multicolumn{2}{|c|}{} & $F_1$ & $F_2$  & $F_1$  & $F_2$& $F_1$  & $F_2$ & $F_1$  &$F_2$ \\
\hline
\multicolumn{2}{|c}{}&\multicolumn{4}{|c}{Homoscedastic lev-1 errors} & \multicolumn{4}{|c|}{Heteroscedastic lev-1 errors}\\
\hline
 \multirow{12}*{\rotatebox[origin=c]{90}{Gaussian}}& $\beta_0$       & 93 & 93.2 & 93 & 93.2 & 91.2 & 92 & 91.2 & 92 \\
&                                                                    & (0.88) & (0.87) & (0.88) & (0.87) & (0.88) & (0.87) & (0.88) & (0.87) \\
& $\beta_1$                                                          & 93.4 & 93.2 & 93.4 & 93.4 & 93.8 & 94.2 & 94 & 94.2 \\
&                                                                    & (0.88) & (0.88) & (0.89) & (0.88) & (0.88) & (0.88) & (0.89) & (0.88) \\
& $\sigma^2_{\epsilon}$                                              & 100 & 0 & 100 & 0.6 & 98.6 & 0.2 & 98.6 & 0 \\
&                                                                    & (0.46) & (0.001) & (0.46) & (0.001) & (0.46) & (0.001) & (0.46) & (0.001) \\
 & $\sigma^2_{u0}$                                                   & 94.4 & 19.6 & 94.4 & 19.8 & 93 & 21.6 & 93 & 22 \\
&                                                                    & (0.73) & (0.09) & (0.73) & (0.09) & (0.73) & (0.09) & (0.73) & (0.09) \\
 &$\sigma_{u01}$                                                     & 92.8 & 23.6 & 92.8 & 23.4 & 92.6 & 27 & 92.8 & 27 \\
&                                                                     & (1.3) & (0.23) & (1.3) & (0.23) & (1.3) & (0.23) & (1.3) & (0.23) \\
 &$\sigma^2_{u1}$                                                     & 93.6 & 24.2 & 94 & 24.8 & 96.4 & 21.2 & 96.8 & 21 \\
&                                                                     & (0.73) & (0.1) & (0.74) & (0.1) & (0.73) & (0.1) & (0.74) & (0.1) \\
\hline
 \multirow{12}*{\rotatebox[origin=c]{90}{$\chi_1^2-1$}}& $\beta_0$         & 91.6 & 91.8 & 91.6 & 91.8 & 92.6 & 91 & 92.8 & 91 \\
&                                                                        & (0.85) & (0.84) & (0.85) & (0.84) & (0.85) & (0.84) & (0.85) & (0.84) \\
&  $\beta_1$                                                             & 91.8 & 89.2 & 91.8 & 89.2 & 91.4 & 90.6 & 91.4 & 90.8 \\
&                                                                        & (0.84) & (0.84) & (0.84) & (0.84) & (0.84) & (0.84) & (0.84) & (0.84) \\
&  $\sigma^2_{\epsilon}$                                                 & 99.6 & 0.2 & 99.6 & 0.2 & 90 & 0 & 90.6 & 0 \\
&                                                                        & (0.56) & (0.001) & (0.56) & (0.001) & (0.56) & (0.001) & (0.56) & (0.001) \\
&   $\sigma^2_{u0}$                                                      & 73 & 8 & 73.2 & 7.4 & 74.6 & 7 & 74.4 & 7 \\
&                                                                        & (0.95) & (0.09) & (0.95) & (0.09) & (0.95) & (0.09) & (0.95) & (0.09) \\
&  $\sigma_{u01}$                                                        & 78 & 18.6 & 78.2 & 19 & 78.6 & 19.4 & 78.6 & 19.4 \\
&                                                                        & (1.44) & (0.21) & (1.45) & (0.21) & (1.44) & (0.21) & (1.45) & (0.21) \\
&  $\sigma^2_{u1}$                                                       & 74.2 & 8 & 74.8 & 7.8 & 78.4 & 9.8 & 78.8 & 9.6 \\
&                                                                        & (0.93) & (0.09) & (0.94) & (0.09) & (0.93) & (0.09) & (0.94) & (0.09) \\
   \hline
\end{tabular}
\end{table}

\normalsize
\section{Conclusions}\label{sec:concl}
In this paper we have investigated the performance of three well established bootstrap schemes for multilevel models: the parametric bootstrap, the residual bootstrap and the cases bootstrap. Also, we have introduced a modified version of the wild bootstrap procedure which is particularly suitable to hierarchical data. We have assessed the finite size performances of the four bootstrap schemes by means of a Monte Carlo study where we have varied sample size, error distribution and error variance. Both the cases bootstrap and the wild bootstrap do not require homoscedasticity and do not make distributional assumptions on the error processes. Still, the performance of the two schemes is very different in terms of coverage and length of confidence intervals. In fact, except for some specific instances, the cases bootstrap has the worst performance of all the four bootstrap schemes, no matter the sample size or the kind of errors. On the contrary, for big sample sizes the wild bootstrap outperforms the three competitors in all the scenarios considered including the Gaussian homoscedastic case. This is especially true as far as estimation of variance components is concerned. In fact, in case of estimation of regression coefficients, both the parametric and the residual bootstrap behave quite well and are robust with respect to non-Gaussianity and heteroscedasticity. The estimation of level-1 variance $\sigma^2_{\epsilon}$, instead, is more problematic: the parametric bootstrap performs very poorly when the assumptions of normality and homoscedasticity are violated; in these cases, the residual bootstrap, behaves better than the parametric bootstrap but it is still outperformed by the wild bootstrap in all the scenarios, with the most dramatic worsening in the heteroscedastic case.
\par
In conclusion, we advocate the use of the proposed version of the wild bootstrap in a multilevel framework especially when the validity of the assumptions underlying the model is questionable, as well as when the sample is sufficiently large.



\end{document}